\documentclass[11pt, letterpaper]{article}
\usepackage{fullpage}
\usepackage{amsmath,amssymb,amsfonts,amsthm,nicefrac}
\usepackage{xspace}
\usepackage{color}
\usepackage{url}
\usepackage{hyperref}
\usepackage{bm}
\usepackage{bbm}
\usepackage{times}
\hypersetup{hidelinks}

\usepackage{enumitem}

\newtheorem{thm}{Theorem}[section]

\newtheorem{rconj}[thm]{Conjecture}
\newtheorem{lemma}[thm]{Lemma}
\newtheorem{corollary}[thm]{Corollary}
\newtheorem{claim}[thm]{Claim}

\newtheorem{definition}[thm]{Definition}
\newtheorem{remark}[thm]{Remark}

\newtheorem{fact}[thm]{Fact}

\newcommand\E{\mathop{\mathbb{E}}}
\newcommand\card[1]{\left| {#1} \right|}
\newcommand\sett[2]{\left\{ \left. #1 \;\right\vert #2 \right\}}

\newcommand\set[1]{{\left\{ #1 \right\}}}
\newcommand\Prob[2]{{\Pr_{#1}\left[ {#2} \right]}}

\newcommand\Expect[2]{{\mathop{\mathbb{E}}_{#1}\left[ {#2} \right]}}

\newcommand\cExpect[3]{{\mathbb{E}_{#1}\left[ \left. #3 \;\right\vert #2 \right]}}
\newcommand\norm[1]{\| #1 \|}

\newcommand\inner[2]{\left\langle{#1},{#2}\right\rangle}
\newcommand\eps{\varepsilon}

\renewcommand\geq{\geqslant}
\renewcommand\leq{\leqslant}

\newcommand{\rom}[1]{\uppercase\expandafter{\romannumeral #1\relax}}

\DeclareMathOperator{\AND}{AND}
\providecommand{\Icrd}{I}
\providecommand{\Iset}{I^{\text{analytic}}}
\providecommand{\mono}{\operatorname{mono}}

\title {On the Proof of Friedgut's DNF Conjecture}

\author{ 
 \textit{Communicated by}\\[0.6em]
    \begin{tabular}{cc}
   Yuval Filmus\thanks{Taub Faculty of Computer Science and Faculty of Mathematics, Technion Israel Institute of Technology. The work was supported by the Israel Science Foundation, Grant No.\ 507/24.}
    &
    \hspace{2em}
    Dor Minzer\thanks{Department of Mathematics, Massachusetts Institute of Technology. Supported by NSF CCF award 2227876 and NSF CAREER award 2239160.}
    \end{tabular}
    }
\date{\vspace{-5ex}}
\begin{document}
\maketitle
\begin{abstract}
Motivated by sharp thresholds in graphs, Friedgut conjectured that  any monotone Boolean function $f\colon \{0,1\}^n\to\{0,1\}$ with total influence at most $K$ with respect to the $p$-biased measure must be $\eps$-close to a width-$w(K,\eps)$ monotone DNF formula, where $w(K,\eps)$ depends only on $K$ and $\eps>0$. This conjecture was recently confirmed by [Liu, Meng], who obtained the width $w(K,\eps) = 2^{O(K^2/\eps^2)}$. 

The main goal of this note is to present a more streamlined proof of Friedgut's conjecture. In doing so, we generalize the statement to general Boolean functions and improve the quantitative value of $w(K,\eps)$ to be optimal. Specifically, we show that any $f\colon \{0,1\}^n\to\{0,1\}$ with total influence at most $K$ with respect to the $p$-biased measure must be $\eps$-close to a \emph{decision list} of width $2^{O(K/\eps)}$. We also obtain versions of Bourgain's tail theorem for the $p$-biased cube.
\end{abstract}
\section{Introduction}
\subsection{Background}
The junta theorem of Friedgut~\cite{FriedgutJunta} asserts that, for the unbiased cube, any function $f\colon \{0,1\}^n\to\{0,1\}$ with small total influence must essentially depend only on a few coordinates. More precisely, defining 
\[
\Icrd_i[f] = \cExpect{x,y\in_R \{0,1\}^n}{x_j=y_j~\forall j\neq i}{\card{f(x)-f(y)}^2},
\qquad
I[f] = \sum\limits_{i=1}^{n}\Icrd_i[f],
\]
Friedgut proved that if $I[f]\leq K$, then for all $\eps>0$, there is a set $J\subseteq [n]$ of size at most $2^{O(K/\eps)}$ and a function $g\colon \{0,1\}^J\to\{0,1\}$ such that
\[
\Prob{x\in_R \{0,1\}^n}{f(x)\neq g(x_J)}\leq \eps.
\]
This result led to a number of important applications, and its underlying philosophy is that local properties of a Boolean function $f$ (not changing its value often when a single coordinate changes) sometimes lead to a global structural characterization of the function. This philosophy is not unique to the cube, and it works in any product space in which the probability of each atom is thought of as a constant.

Due to applications to sharp thresholds, Friedgut~\cite{Friedgut} proposed that a similar philosophy should hold for the $p$-biased cube for any $p\in (0,1)$. Here and throughout, we will fix the convention that $p\leq 1/2$,  define the $p$-biased measure as $\mu_p^{n}(x) = p^{\card{x}}(1-p)^{n-\card{x}}$, and the corresponding notion of influences by
\[
\Icrd_i[f;\mu_p] = \Expect{\substack{x\sim\mu_p^{n},y_i\sim \mu_p,y_{-i}=x_{-i}}}{\card{f(x)-f(y)}^2},
\qquad
I[f;\mu_p] = \sum\limits_{i=1}^{n}\Icrd_i[f;\mu_p].
\]
It turns out that the statement of Friedgut's theorem over the unbiased cube fails over the $p$-biased cube, as for example one can show that the function $f(x) = x_1\lor \dots\lor x_{1/p}$ has constant variance and $I[f;\mu_p]=O(1)$, yet it is not close to any $O(1)$-junta. Motivated by this example, Friedgut made the following conjecture. Here and throughout, a function $f\colon \{0,1\}^n\to\{0,1\}$ is called \emph{monotone} if $f(x)\leq f(y)$ whenever $x_i\leq y_i$ for all $i$. A DNF formula is \emph{monotone} if it only involves positive literals.
\begin{rconj}\label{conj:friedgut}
    Let $K\geq 1$. If $f\colon \{0,1\}^n\to\{0,1\}$ is monotone and $I[f;\mu_p]\leq K$, then for all $\eps>0$, $f$ is $\eps$-close to a monotone DNF formula of width $w(K,\eps)$.
\end{rconj}
In his paper proposing this conjecture, Friedgut managed to prove it for functions $f$ that exhibit sufficient symmetry, such as graph and hypergraph properties. His paper contains an appendix Bourgain, which establishes a more modest \emph{booster result} for general monotone functions. Here, a booster is a set of coordinates $B\subseteq [n]$ of size $O_{K,\eps}(1)$ such that
\[
\Pr_{x\sim \mu_p^{n}}\left[f(x)=1 \;\middle|\; x_B=\vec{1}\right]
\geq \Prob{x\sim \mu_p^{n}}{f(x)=1}+\Omega_{K}(1),
\]
and Bourgain's result asserts that such a booster exists as long as $I[f]\leq K$ and $0 \ll \E[f] \ll 1$. Hatami~\cite{Hatami} made further progress on Conjecture~\ref{conj:friedgut} and showed that any such function must be close to a notion he refers to as a ``pseudo-junta''. Building on Hatami's argument, Liu and Meng~\cite{LM26} recently managed to resolve Conjecture~\ref{conj:friedgut} and obtained $w(K,\eps) = 2^{O(K^2/\eps^2)}$.

\subsection{Main Result}
The original motivation for this work stemmed from trying to understand the arguments of~\cite{Hatami,LM26} and find a more natural way to view them. In doing so, we managed to first of all show that Conjecture~\ref{conj:friedgut} holds with the optimal bound of $w(K,\eps) = 2^{O(K/\eps)}$, second that there is a statement that generalizes it to functions $f$ that are not necessarily monotone, and third that one can strengthen it along the lines of Bourgain's tail theorem~\cite{Bourgain,KKO}. The correct generalization of DNF formulas turns out to be decision lists, as defined below.
\begin{definition}\label{def:decision_list}
    A \emph{(positive) decision list}\footnote{General decision lists allow rules of the form $x|_{S_i} = y_i$ for arbitrary $y_i \in \{0,1\}^{S_i}$. We focus on positive decision lists since they generalize monotone DNFs.} of width $w$ is an ordered sequence of set-bit pairs $(S_1,b_1),\ldots,$ $(S_N,b_N)$, also called \emph{rules}, and a bit $b_{N+1}$, where $\card{S_i}\leq w$ for all $i$. The Boolean function computed by the decision list is $g\colon \{0,1\}^n\to\{0,1\}$ defined by $g(x) = b_i$ for the first $i$ such that $\bigwedge_{j\in S_i}x_j=1$ (where we take $i=N+1$ if no such $i$ exists).
\end{definition}
A monotone DNF formula of width $w$ with $N$ terms can be viewed as a width $w$ positive decision list with $b_1=\dots=b_N=1$ and $b_{N+1}=0$, so the class of positive decision lists generalizes monotone DNF formulas. Next, we also note that if $g$ is a positive decision list of width $w$, then $I[g;\mu_p]\leq O(w)$ (see Fact~\ref{fact:bound_ttf}), so small-width positive decision lists provide natural examples of functions with small total influence. The main result of this note is that the converse is (essentially) true.
\begin{thm}\label{thm:main}
    Let $K\geq 1$. Suppose $f\colon (\{0,1\}^n,\mu_p^{n})\to\{0,1\}$ is a function with $I[f; \mu_p]\leq K$. Then for all $\eps>0$, there is a function $g$ that can be computed by a positive decision list of width $2^{O(K/\eps)}$, such that
    \[
    \Prob{x\sim\mu_p^{n}}{f(x)\neq g(x)}\leq \eps.
    \]
\end{thm}
With some adaptations to the proof, we get the following result for the monotone case.
\begin{thm}\label{thm:main_mono}
    Let $K\geq 1$. Suppose $f\colon (\{0,1\}^n,\mu_p^{n})\to\{0,1\}$ is a monotone function with $I[f; \mu_p]\leq K$. Then for all $\eps>0$, there is a function $g$ that can be computed by a monotone DNF of width $2^{O(K/\eps)}$, such that
    \[
    \Prob{x\sim\mu_p^{n}}{f(x)\neq g(x)}\leq \eps.
    \]
\end{thm}
We remark that Theorems~\ref{thm:main},~\ref{thm:main_mono} are asymptotically tight with respect to the width, and in Section~\ref{sec:tightness} we show matching examples.
\begin{remark}
   One can obtain a similar result to Theorem~\ref{thm:main} over more general product domains, as well as for product-like domains such as the symmetric group. In these results, the rules of the decision lists are functions of the form $\prod\limits_{i\in W}1_{x_i\in Q_i}$, which is thought of as the analog of an AND function of width $\card{W}$. One can prove these result by direct reductions to the statements about the $p$-biased cube proved herein, or alternatively modify the proof given herein to these settings. For the latter, one has to use the appropriate version of global hypercontractivity in each setting (which is due to~\cite{KLLMcodes} for product domains and~\cite{keevash2023sharp} for the symmetric group). We omit the full details.
\end{remark}
\subsubsection{Tail Theorems}
Our next result is an analog of the classical tail theorem of Bourgain~\cite{Bourgain,KKO}, which asserts that a Boolean function $f$ over the unbiased Boolean hypercube satisfying $W^{\geq k}[f]\leq \frac{\eps}{\sqrt{k}}$ must be $O(\eps)$-close to a junta of size $\frac{2^{O(k)}}{\eps^{O(1)}}$. Here and throughout, $W^{\geq k}[f] = \sum\limits_{\card{S}\geq k}\widehat{f}(S)^2$ is the Fourier mass that $f$ has on level $k$ onwards. Noting that $W^{\geq k}[f]\leq \frac{I[f]}{k}$ and taking $k = (I[f]/\eps)^2$, one can observe that this result is a qualitative strengthening of Friedgut's junta theorem~\cite{FriedgutJunta} (however it gives worse quantitative bounds). The following results can be seen as an analogs of Bourgain's result to the $p$-biased cube.
\begin{thm}\label{thm:Bourgain}
    Suppose $f\colon (\{0,1\}^n,\mu_p^n)\to\{0,1\}$ is a function with $W^{\geq k}[f]\leq \frac{\eps}{\sqrt{k}}$. Then there exists a function $g$ that can be computed by a positive decision list of width $2^{O(k)}/\eps^{O(1)}$ such that
    \[
    \Prob{x\sim \mu_p^n}{f(x)\neq g(x)}\leq O(\eps).
    \]
\end{thm}

For monotone functions, the structure of positive decision lists can be replaced once again by monotone DNF formula, as follows.
\begin{thm}\label{thm:Bourgain_mono}
    Suppose $f\colon (\{0,1\}^n,\mu_p^n)\to\{0,1\}$ is a monotone function with $W^{\geq k}[f]\leq \frac{\eps}{\sqrt{k}}$. Then there exists a function $g$ that can be computed by a monotone DNF formula of width $2^{O(k)}/\eps^{O(1)}$ such that
    \[
    \Prob{x\sim \mu_p^n}{f(x)\neq g(x)}\leq O(\eps).
    \]
\end{thm}

\subsection{Proof Overview}
In this section we outline the proof of Theorem~\ref{thm:main}, and compare it to the proofs in~\cite{Friedgut,Hatami}. Fix a function $f\colon \{0,1\}^n\to\{0,1\}$ with $I[f;\mu_p]\leq K$ henceforth, and write its $p$-biased Fourier expansion
\[
f(x) = \sum\limits_{S\subseteq[n]}\widehat{f}(S)\chi_S^{p}(x),
\]
where $\chi_S^{p}(x) = \prod\limits_{i\in S}\frac{x_i-p}{\sqrt{p(1-p)}}$ are the $p$-biased characters and $\widehat{f}(S) = \langle f,\chi_S^{p}\rangle$ are the Fourier coefficients of $f$, where we use the standard $L_2$ inner product $\langle f,g\rangle = \Expect{x\sim\mu_p^n}{f(x)g(x)}$. 

\subsubsection{The Main Argument} The proof consists of the following steps:
\begin{enumerate}
    \item \textbf{Low-degree part:} letting $d= \frac{K}{\eps}$, we first argue that all but $\eps$ of the Fourier mass of $f$ lies below level $d$. This is a standard consequence of the fact that the total influence $I[f;\mu_p]$ of a function is equal to its average degree $2\sum\limits_{S}\card{S}\widehat{f}(S)^2$. Henceforth, we take $g$ to be the part of $f$ of degree at most $d$, i.e., $g(x)=\sum\limits_{\card{S}\leq d}\widehat{f}(S)\chi_S(x)$.
    \item \textbf{Influential coalitions and cleaning up the Fourier expansion:} we next  define the notion of influential subsets of variables, which generalizes the notion of influential coordinates. We then remove from the Fourier transform of $f$ any character $S$ that cannot be written as a union of influential coalitions. 
    
    More precisely, for each $i\in [n]$, we define the discrete derivative $\partial_i f\colon \{0,1\}^n\to \mathbb{R}$ by 
    \[
    \partial_if(y) = f(x_{-i}=y_{-i},x_i=1)-f(x_{-i}=y_{-i},x_i=0),
    \]
    and then define the derivative with respect to a set $T = \{i_1,\ldots,i_s\}$ as $\partial_T f = \partial_{i_1}\partial_{i_2}\ldots\partial_{i_s}f$ (the order turns out not to matter). It is easy to see that, for a Boolean function $f$, the influence of variable $i$ is related to the norm of the derivative with respect to $i$ as $\Icrd_i[f] = 2p(1-p)\norm{\partial_i f}_2^2$, and generalizing this we define the influence of a set $T$ to be $\Iset_T[f] = \norm{\partial_T f}_2^2$ (note that the normalization is different, and in particular, $\Icrd_i[f] = 2p(1-p)\Iset_{\{i\}}[f]$).
    Taking a parameter $\eta = 2^{-C\cdot \frac{K}{\eps}}$ for a large constant $C>0$ to be determined, we define the collection of influential coalitions as 
    $\mathcal{H} = \sett{T\neq \emptyset}{\Iset_T[g]\geq \eta}$, and then define 
    \[
    \mathcal{S} = \left\{S\subseteq [n] \;\middle|\; \card{S}\leq d, S = \bigcup_{T\in \mathcal{H}, T\subseteq S}T\right\}
    \]
    and $h(x) = \sum\limits_{S\in\mathcal{S}}\widehat{g}(S)\chi_S(x)$. We show that $g$ and $h$ are close in $L_2$-norm provided $C$ is sufficiently large, and this is done using global hypercontractivity~\cite{KLLM}.
    \item \textbf{Moving from Fourier basis to AND basis:} 
    note that neither the set of influential coalitions $\mathcal{H}$ nor the set of characters $\mathcal{S}$ need to be sparse; indeed both could be of size as large as $p^{-\Omega(d)}$. However, for $x\sim \mu_p^{n}$, the expected number of $T\in\mathcal{H}$ such that $\bigwedge_{i\in T} x_i=1$ is equal to $\sum\limits_{T\in\mathcal{H}}p^{|T|}\leq \frac{1}{\eta}\sum\limits_{T}p^{|T|}\Iset_T[g]$, which can be shown to be $\frac{2^{O(d)}}{\eta}$ by expressing $\Iset_T[g]$ using the Fourier coefficients of $g$. Thus, on average, if we take the AND basis with respect to $\mathcal{H}$, the number of ANDs that fire will typically be very small, and furthermore, each character in $h$ is ``covered'' by sets from $\mathcal{H}$. This suggests that if we move in $h$ from the Fourier basis to the AND basis, then we will have a representation which is ``effectively sparse''. This is precisely the goal of this step.

    Formally speaking, for each $S\subseteq [n]$ and $y\in\{0,1\}^S$ we define the function $\chi_{S,y}(x) = \prod\limits_{i\in S}\frac{x_i-y_i}{\sqrt{p(1-p)}}$. On the one hand, clearly we have that 
    \begin{equation}\label{eq:pseudo_character_to_and}
    \chi_{S,y}(x) = \frac{1}{\left(\sqrt{p(1-p)}\right)^{|S|}} (-1)^{|y|}\bigwedge_{i\in S}1_{x_i=1-y_i},
    \end{equation}
    so $\chi_{S,y}$ is a multiple of an AND function over $S$, which can therefore be written as a product of AND functions over some $T\in \mathcal{H}$. 
    On the other hand, averaging over $y$ we see that $\chi_S(x) = \Expect{y\sim\mu_p^{S}}{\chi_{S,y}(x)}$, so we may hope that $\chi_S(x)$ can be approximated well by taking a few random $y$'s and averaging the functions $\chi_{S,y}$. In that case, in the function $h$ the characters $\chi_S$ can potentially be replaced by linear combinations of $\chi_{S,y(i)}$.

    This hope turns out to be correct. Sampling $y(1),\ldots,y(m)\sim\mu_p^{n}$ independently for not-too-large $m$, one may show that the functions $x\rightarrow \chi_{S}(x)$ and $x\rightarrow \Expect{i\in[m]}{\chi_{S,y(i)}(x)}$ are typically close in $L_2$ distance. Taking
    \[
    h_{\vec{y}}(x)
    =\sum\limits_{S\in\mathcal{S}}\widehat{h}(S)\Expect{j\in [m]}{\chi_{S,y(j)}(x)},
    \]
     a direct computation gives that $\Expect{\vec{y}}{\norm{h_{\vec{y}}-h}_2^2}\leq \frac{2^d}{m}$. We now build the collection of AND functions
    \[
    \mathcal{C}_{\vec{y}}=\left\{\prod\limits_{i\in T}1_{x_i = 1-y(j)_i}\right\}_{T\in\mathcal{H}, j=1,\ldots,m},
    \]
    and notice that $h_{\vec{y}}$ is a linear combination of products of AND functions from the collection $\mathcal{C}_{\vec{y}}$. Indeed, combining~\eqref{eq:pseudo_character_to_and} and the fact that each $S\in\mathcal{S}$ is covered by $\mathcal{H}$, we are able to replace $\chi_{S,y(j)}(x)$ by a product of functions from $\mathcal{C}$.
    Combining the results thus far, with high probability over the choice of $\vec{y}$ we get that  $f$ and $h_{\vec{y}}$ are close in $L_2$.
    \item \textbf{Getting the Decision List:} let us think of $\mathcal{C}_{\vec{y}}$ as a collection of functions $A_1,\ldots,A_N$, each being an AND function over some set $T\in\mathcal{H}$. Then we know that $h_{\vec{y}}(x)$ is only a function of $(A_1(x),\ldots,A_N(x))$, so for each $I\subseteq [N]$ it is constant on each set of the form
    \[
    E_I = \sett{x}{A_i(x)=1\iff i\in I}.
    \]
    Here, one should think of the sets $E_I$ as partitioning the space according to which of the ANDs $A_i$ are $1$. Since $h_{\vec{y}}$ is constant on $E_I$, we may choose a Boolean value $b(I)\in\{0,1\}$ closest to $h_{\vec{y}}$ on $E_I$. This gives us a (non-positive) decision list whose terms are $A_I(x) = \prod\limits_{i\in I}A_i(x)$ with value $b(I)$, ordered by decreasing size of $\card{I}$.
    This decision list is not of bounded width though, and to repair that we note that for a typical $x$, at most $r = m2^{O(d)}$ of the $A_i(x)$ are $1$.\footnote{This fact is true with high probability over the choice of $\vec{y}$, but we ignore this for simplicity.} Thus, if we only keep rules with $\card{I}\leq r/\eps$, the computed function does not change by much, hence it approximates $h_{\vec{y}}$ and therefore approximates $f$.
    
    The slight issue with this argument is that it produces a decision list with terms containing negated variables, and to circumvent this more care is needed. Let us write $A_i(x) = A_{i,+}(x)\cdot A_{i,-}(x)$, where $A_{i,+}(x)$, $A_{i,-}(x)$ are the parts of the AND containing non-negated and negated variables, respectively, let $V_i$ be the set of variables in $A_i$, and let $V_{i,+}$ be the set of variables in $A_{i,+}$. 
    For each $I\subseteq [N]$ define
    \[
    A_I(x) = \prod\limits_{i\in I}A_{i,+}(x),
    \qquad
    V_{I,+} = \bigcup_{i\in I} V_{i,+},
    \qquad
    V_{I} = \bigcup_{i\in I} V_{i}.
    \]
    Ideally, we would have liked to say that $h_{\vec{y}}$ is constant on the analog of the set $E_I$ from before where we only consider the positive parts $A_{i,+}$, alas this is not true because it does not give us any information on the assignment to the variables in $V_{I}\setminus V_{I,+}$. To fix that, refine further, and for each $V_{I,+}\subseteq P\subseteq V_I$ we take
    \[
    E_{I,P} = \sett{x}{A_{i,+}(x)=1\iff i\in I, \;\;\forall j\in V_I, x_j=1\iff j\in P}.
    \]
    It is not hard to see that the partition $\{E_{I,P}\}_{I,P}$ is a refinement of the previous partition $\{E_I\}_{I}$, and thus $h_{\vec{y}}$ is constant on each $E_{I,P}$. To build a decision list we take $b(I,P)\in\{0,1\}$ to be Boolean value closest to $h_{\vec{y}}$ on $E_{I,P}$, and define the rules of our decision list to be $(P,b(I,P))$ ordered by the decreasing lexicographical ordering on $(\card{I},\card{P})$.\footnote{I.e., the tuples with larger $\card{I}$ appear first, and among the ones with the same $\card{I}$, those with larger $\card{P}$ appear first.} This decision list exactly computes the rounding of $h_{\vec{y}}$ to a Boolean function. As before, to turn it into a bounded width decision list, we again observe that we may discard rules related to $I$ or $P$ that are too large, and only introduce a small amount of error.
\end{enumerate}

\subsubsection{Comparison of Proof to Prior Works}
We compare each one of these steps to analogous steps in prior works:
\begin{enumerate}
    \item \textbf{Low-degree part:} this step appears in many of the related works~\cite{KKL,FriedgutJunta,Friedgut,Hatami} in exactly the same way.
    \item \textbf{Influential coalitions and cleaning up the Fourier expansion:} analogs of this part of the argument appear in many of the related works. In the unbiased cube, it first appeared in the work~\cite{KKL} and more clearly in~\cite{Friedgut}, where the authors bound the total Fourier mass of low-degree characters that contain some small influence variable. In the biased cube, the work of~\cite{Friedgut} makes a more ad-hoc argument based on the symmetries of the function $f$, and shows that the total contribution of ``non-modest characters $S$'' to the Fourier spectrum is negligible. In the appendix of~\cite{Friedgut}, Bourgain gives a more general way to clean up the Fourier transform of $f$ of any $S$ that are not covered by influential coalitions. His method works for more general functions $f$ but obtains worse quantitative bounds (a more accessible description of Bourgain's method can be found in~\cite{ODonnell}), and the work~\cite{Hatami} makes a direct use of Bourgain's argument.
    \item \textbf{Moving from the Fourier basis to the AND basis:} this part is not needed in the unbiased cube, since the functions already become Fourier sparse after the previous clean-up process. Over the biased cube, the work~\cite{Friedgut} effectively moved to the AND basis using symmetries of the function. More precisely, in the context therein there are only $O_{d}(1)$ many distinct orbits of characters $S$, and the Fourier coefficients in each orbit are the same. Thus, the function $h$ can be written as a sum of $O_{d}(1)$ many functions, each being a multiple of a simple sum of characters, and simple sum of characters can be directly translated to a suitable AND basis. The work~\cite{Hatami} was unable to directly move to an AND basis, and instead settled for the weaker structure of pseudo-junta. It is basically the best approximating function $g$ of $f$, in which for each $T\in\mathcal{H}$ we have a function $J_T\colon \{0,1\}^T\to\{0,1\}$, and $g(x)$ only depends on $x_T$ for all $T$ such that $J_T(x_T)=1$.
    \item \textbf{Getting the Decision List:} the proof of~\cite{LM26} contains a component that is close in spirit to this step. In their argument they first take the pseudo-junta produced by Hatami's result and then show that it can be turned into a monotone pseudo-junta (somewhat parallel to the way we have removed negated variables in the decision list). They then show how to approximate a monotone pseudo-junta by a monotone DNF formula.
\end{enumerate}

\paragraph{Statement of AI use:} 
we prompted ChatGPT 6 to use global hypercontractivity to improve the pseudo-junta result of~\cite{Hatami} to $2^{O(K)}$, and then use it to improve the result of~\cite{LM26}. It initially failed, but after some encouragement it succeeded. Studying the proof and simplifying it, we realized that a plausible candidate structure for non-monotone functions is decision lists, which ChatGPT 6 quickly confirmed to be correct. We also realized a similar-in-spirit argument may work to obtain a version of Bourgain's tail theorem in the $p$-biased cube, and ChatGPT 6 once again quickly confirmed it.
The current manuscript was rewritten by the communicators in an effort to present these proofs in a human-friendly manner. 

\section{Preliminaries}
\subsection{Basics of Fourier Analysis}
In this section we collect a few basic facts from Fourier analysis. Throughout, we will think of the Boolean cube $\{0,1\}^n$ equipped with the $p$-biased measure, and consider the standard $L_2$ inner product on real-valued functions over it. We begin with the Parseval/Plancherel identity:
\begin{fact}\label{fact:parseval}
    Suppose $f,g\colon \{0,1\}^n\to\mathbb{R}$ are functions. Then $\inner{f}{g}=\sum\limits_{S}\widehat{f}(S)\widehat{g}(S)$.
\end{fact}
\begin{proof}
    Immediate from the orthogonality of the $p$-biased characters.
\end{proof}
Next, we give Fourier-analytic formulas for the influences:
\begin{fact}\label{fact:influence}
    Suppose $f\colon \{0,1\}^n\to\mathbb{R}$ is a function. Then for each $i\in [n]$ we have that 
    \begin{enumerate}
        \item $\Icrd_i[f;\mu_p]=2\sum\limits_{S\ni i}\widehat{f}(S)^2$.
        \item $I[f;\mu_p]=2\sum\limits_{S}\card{S}\widehat{f}(S)^2$.
        \item $\partial_if(x) = \frac{1}{\sqrt{p(1-p)}}\sum\limits_{S\ni i}\widehat{f}(S)\chi_{S\setminus\{i\}}(x)$.
    \end{enumerate}
\end{fact}
\begin{proof}
    Expanding $f$ into Fourier transform, we get that sampling $x\sim\mu_p^{n}$ and then $y\sim\mu_p^n$ conditioned on $y_j = x_j$ for all $j\neq i$, we have that
    \[
    f(x)-f(y)=\sum\limits_{S}\widehat{f}(S)\chi_{S\setminus\{i\}}(x)(\chi_{i}(x)-\chi_{i}(y)),
    \]
    so only $S$ such that $i\in S$ have non-zero contribution. Squaring, expanding and taking expectation gives that 
    \[
    \Icrd_i[f;\mu_p]=
    \Expect{x,y}{(f(x)-f(y))^2}
    =\sum\limits_{S\ni i}\widehat{f}(S)^2\Expect{x_i,y_i}{(\chi_{i}(x_i)-\chi_{i}(y_i))^2}
    =2\sum\limits_{S\ni i}\widehat{f}(S)^2,
    \]
    giving the first item. The second item follows by summing over all $i$, and the third item follows by plugging in $x_i=1$, $y_i=0$ and simplifying.
\end{proof}
Finally, we give Fourier-analytic formulas for the generalized influences.
\begin{fact}\label{fact:generalized_influences}
    If $f\colon \{0,1\}^n\to\mathbb{R}$ is a function and $T\subseteq [n]$ is any set, then 
    $\Iset_T[f]=\frac{1}{(p(1-p))^{\card{T}}}\sum\limits_{S\supseteq T}\widehat{f}(S)^2$.
\end{fact}
\begin{proof}
    Using the third item in Fact~\ref{fact:influence} we get that $\partial_T f(x) = \frac{1}{\sqrt{p(1-p)}^{\card{T}}}\sum\limits_{S\supseteq T}{\widehat{f}(S)\chi_{S\setminus T}(x)}$, and the result follows from Fact~\ref{fact:parseval}.
\end{proof}

\subsection{Global Hypercontractivity}
We need the following global hypercontractivity result from~\cite[Lemma 3.6]{KLLM}.
\begin{lemma}\label{lem:GH}
    Suppose $f\colon \{0,1\}^n\to\mathbb{R}$ is a function of degree $d$, and that $\Iset_T[f]\leq \eta$ for all $T\subseteq [n]$. Then
    \[
    \norm{f}_4\leq 5^d\eta^{1/4}\norm{f}_2^{1/2}.
    \]
\end{lemma}

\section{Proof of Theorem~\ref{thm:main}}\label{sec:proof_of_main}
In this section we prove Theorem~\ref{thm:main}. We fix a function $f$ as in the statement henceforth.

\subsection{The First Step: Focusing on Low-degrees}
Let $d=\frac{K}{\eps}$ and define $g(x) = \sum\limits_{\card{S}\leq d}\widehat{f}(S)\chi_S(x)$. Then by Fact~\ref{fact:parseval} and Fact~\ref{fact:influence}
\begin{equation}\label{eq:remove_high_eg}
    \norm{f-g}_2^2
    =\sum\limits_{\card{S}>d}\widehat{f}(S)^2
    \leq \sum\limits_{S}\frac{\card{S}}{d}\widehat{f}(S)^2
    =\frac{I[f]/2}{d}\leq \frac{\eps}{2},
\end{equation}
and we henceforth work with the function $g$. 
\subsection{The Second Step: Cleaning-up Low-influence Coalitions}
Let $\eta = 2^{-20\cdot d}$, and define
\[
\mathcal{H} = \sett{T\neq \emptyset}{\Iset_T[g]\geq \eta}.
\]
Also, define the set of characters that $\mathcal{H}$ can form by unions as
\[
\mathcal{S} = \sett{S\subseteq [n]}{\exists T_1,\ldots,T_q\in\mathcal{H}, T_1\cup\dots \cup T_q = S},
\]
and define $h(x) = \sum\limits_{S\in\mathcal{S}}\widehat{g}(S)\chi_S(x)$. The following claim shows that $g$ and $h$ are close in $L_2$.
\begin{claim}\label{claim:g_h_close}
    $\norm{g-h}_2^2\leq \frac{1}{2}5^d\eta^{1/4} K\leq \eps$.
\end{claim}
\begin{proof}
    By Fact~\ref{fact:parseval} we have that $\norm{g-h}_2^2 = \sum\limits_{\card{S}\leq d, S\not\in\mathcal{S}}\widehat{f}(S)^2$, and we next split this sum by coordinates. For $i\in [n]$ and $S\subseteq[n]$ containing $i$, we say $i$ is bad for $S$ if for all $T$ such that $i\in T\subseteq S$, we have that $T\not\in \mathcal{H}$. Note that if $S\not\in\mathcal{S}$, then some $i\in S$ is bad for $S$. Indeed, otherwise for each $i$ we would have some $T_i\in\mathcal{H}$ such that $i\in T_i\subseteq S$, and then $S = \bigcup_{i\in S} T_i$ and so $S\in\mathcal{S}$. Thus, defining $\mathcal{B}_i = \sett{S}{\card{S}\leq d, i\text{ the smallest bad element for $S$}}$ we get that 
    \[
    \norm{g-h}_2^2
    = \sum\limits_{i=1}^n\sum\limits_{S\in \mathcal{B}_i}\widehat{f}(S)^2.
    \]
    Writing $\xi_i(x) = \frac{1}{\sqrt{p(1-p)}}\sum\limits_{S\in\mathcal{B}_i}\widehat{f}(S)\chi_{S\setminus \{i\}}(x)$, we get from the above and Facts~\ref{fact:parseval},~\ref{fact:influence} that 
    \[
    \norm{g-h}_2^2
    = p(1-p)\sum\limits_{i=1}^{n}\inner{\xi_i}{\partial_if}
    \leq p(1-p)\sum\limits_{i=1}^{n}\norm{\xi_i}_4\norm{\partial_if}_{4/3},
    \]
    where in the last transition we used H\"older's inequality. Since $\partial_i f$ is $\{-1,0,1\}$ valued we have that $\norm{\partial_i f}_{4/3} = \Prob{x}{\partial_i f(x)\neq 0}^{3/4}$. For $\xi_i$, note that by definition if $T\cup\{i\}\not\in \mathcal{H}$ then $\Iset_T[\xi_i]\leq \Iset_{T\cup\{i\}}[g]\leq \eta$, and else $\partial_T \xi_i\equiv 0$ so $\Iset_T[\xi_i]=0$. In either case, $\Iset_T[\xi_i]\leq \eta$ for all $T$, so by Lemma~\ref{lem:GH} we get that 
    \begin{equation}\label{eq:global_hyp}
    \norm{\xi_i}_4\leq 5^d\eta^{1/4}\norm{\xi_i}_2^{1/2}\leq 5^d\eta^{1/4}\norm{\partial_i f}_2^{1/2}
    =5^d\eta^{1/4}\Prob{x}{\partial_i f(x)\neq 0}^{1/4},
    \end{equation}
    where we used Fact~\ref{fact:parseval} in the second transition. Combining, we get that 
    \[
    \norm{g-h}_2^2
    \leq p(1-p)
    5^d\eta^{1/4}\sum\limits_{i=1}^n\Prob{x}{\partial_i f(x)\neq 0}
    =5^d\eta^{1/4}\frac{1}{2}\sum\limits_{i=1}^n \Icrd_i[f]
    \leq \frac{1}{2}5^d\eta^{1/4} K.
    \qedhere
    \]
\end{proof}
\subsection{The Third Step: Moving to the AND Basis}
Set $m = \frac{2^d}{\eps}$, sample $y(1),\ldots,y(m)\sim\mu_p^n$ independently, and define the function
\[
h_{\vec{y}}(x)=\sum\limits_{S\in\mathcal{S}}\widehat{g}(S)\Expect{j\in[m]}{\chi_{S,y(j)}(x)},
\]
where we recall from the introduction that $\chi_{S,y}(x)=\prod\limits_{i\in S}\frac{x_i-y_i}{\sqrt{p(1-p)}}$.
\begin{claim}\label{claim:hy_close}
    $\Expect{\vec{y}}{\norm{h_{\vec{y}}-h}_2^2}\leq \frac{2^d-1}{m}\norm{g}_2^2\leq \eps$.
\end{claim}
\begin{proof}
   Expanding, we have that 
   \[
   \Expect{\vec{y}}{\norm{h_{\vec{y}}-h}_2^2}
   =\sum\limits_{S,S'}\widehat{g}(S)\widehat{g}(S')
   \Expect{x,\vec{y}}{\left(\chi_S(x)-\Expect{j\in[m]}{\chi_{S,y(j)}(x))}\right)
   \left(\chi_{S'}(x)-\Expect{j\in[m]}{\chi_{S',y(j)}(x)}\right)}.
   \]
   Fix $S\neq S'$, and suppose without loss of generality that $i\in S\setminus S'$. Then we may push the expectation over $x_i,y_i$ inside on the first term (as the second one is independent of it), and note that as $\Expect{x_i}{\chi_{\{i\}}(x)}=\Expect{x_i,\vec{y}_i}{\chi_{\{i\},y(j)}(x)}=0$ we get that it vanishes. Thus, the contribution of $S\neq S'$ to the right hand side vanishes, so
   \begin{align*}
   \Expect{\vec{y}}{\norm{h_{\vec{y}}-h}_2^2}
   &=\sum\limits_{S}\widehat{g}(S)^2
   \Expect{x,\vec{y}}{\left(\chi_S(x)-\Expect{j\in[m]}{\chi_{S,y(j)}(x)}\right)^2}\\
   &=\sum\limits_{S}\widehat{g}(S)^2
   \left(\Expect{x,\vec{y}}{\left(\Expect{j\in[m]}{\chi_{S,y(j)}(x)}\right)^2-1}\right).
   \end{align*}
   Expanding,
   \begin{align*}
   \Expect{x,\vec{y}}{\left(\Expect{j\in[m]}{\chi_{S,y(j)}(x)}\right)^2}
   &=\frac{1}{m^2}
   \sum\limits_{j=1}^m
   \Expect{x,\vec{y}}{\chi_{S,y(j)}(x)^2}
   +\frac{1}{m^2}\sum\limits_{j\neq j'}
   \Expect{x,\vec{y}}{\chi_{S,y(j)}(x)\chi_{S,y(j')}(x)}\\
   &=\frac{1}{m^2}
   \sum\limits_{j=1}^m
   \Expect{x,\vec{y}}{\chi_{S,y(j)}(x)^2}
   +\frac{1}{m^2}\sum\limits_{j\neq j'}
   \Expect{x}{\chi_{S}(x)^2},
   \end{align*}
   where we used $\Expect{y(j)}{\chi_{S,y(j)}(x)}=\chi_S(x)$. 
   We have
   \[
   \Expect{x,\vec{y}}{\chi_{S,y(j)}(x)^2}=\prod\limits_{i\in S}2p(1-p)\frac{(1-0)^2}{p(1-p)}= 2^{\card{S}}
   \leq 2^d
   \]
   and $\Expect{x}{\chi_S(x)^2}=1$,  so we conclude that
   \[
   \Expect{x,\vec{y}}{\left(\Expect{j\in[m]}{\chi_{S,y(j)}(x)}\right)^2}
   \leq \frac{2^d}{m}+\frac{m-1}{m}.
   \]
   Plugging this above gives
   $\Expect{\vec{y}}{\norm{h_{\vec{y}}-h}_2^2}\leq 
   \sum\limits_{S}\widehat{g}(S)^2\cdot\frac{2^d-1}{m}$, giving the claim.
\end{proof}
Next, we introduce the collection of AND functions defined by $y(1),\ldots,y(m)$, and their positive parts. For each $T\in\mathcal{H}$ and $j\in [m]$, define 
the function $\AND_{T,j}(x)=\prod\limits_{i\in T}1_{x_i=1-y(j)_i}$, and split it as
\[
\AND_{T,j}(x)
=\AND_{T,j,+}(x)
\cdot 
\AND_{T,j,-}(x),
\]
where $\AND_{T,j,+}$ consists of the product over coordinates $i$ where $y(j)_i=0$ and $\AND_{T,j,-}$ consists over the product over coordinates $i$ where $y(j)_i=1$.
\begin{claim}\label{claim:number_of_ANDs_firing}
    $\Expect{\vec{y}}{\sum\limits_{T\in\mathcal{H}}\sum\limits_{j=1}^{m}\Expect{x}{\AND_{T,j,+}(x)}}\leq \frac{m4^d}{\eta}$.
\end{claim}
\begin{proof}
    Pushing the expectation over $y$ inside, we get that 
    \[
    \Expect{x,\vec{y}}{\AND_{T,j,+}(x)}
    =(p+p(1-p))^{\card{T}},
    \]
    so the left hand side of the claim is equal to
    \[
    m\sum\limits_{T\in\mathcal{H}}(p+p(1-p))^{\card{T}}
    \leq \frac{m}{\eta}\sum\limits_{T}(p+p(1-p))^{\card{T}}\Iset_T[g]
    =\frac{m}{\eta}\sum\limits_{T}\sum\limits_{S\supseteq T}\left(\frac{2-p}{1-p}\right)^{\card{T}}\widehat{g}(S)^2,
    \]
    where in the last transition we used Fact~\ref{fact:generalized_influences}. Since $p\leq 1/2$, we have that $\frac{2-p}{1-p}\leq 3$, and the last sum is at most $\sum\limits_{S}4^{\card{S}}\widehat{g}(S)^2\leq 4^{d}$, and the claim follows.
\end{proof}

Combining Claims~\ref{claim:hy_close},~\ref{claim:number_of_ANDs_firing} with Markov's inequality and the union bound, we may find $\vec{y}$ such that 
\begin{equation}\label{eq:unionbd}
    \norm{h_{\vec{y}}-h}_2^2\leq 3\eps,
    \qquad \qquad
    \sum\limits_{T\in\mathcal{H}}\sum\limits_{j=1}^{m}\Expect{x}{\AND_{T,j,+}(x)}\leq 3\frac{m4^d}{\eta},
\end{equation}
and we fix it henceforth.
\subsection{The Fourth Step: Creating the Decision List}\label{sec:4th_step}
For each $I\subseteq\mathcal{H}\times [m]$, define $V_{I}$ to be the set of variables appearing in $\{\AND_{T,j}~|(T,j)\in I\}$, and define $V_{I,+}$ to be the set of variables appearing in $\{\AND_{T,j,+}~|(T,j)\in I\}$. 
For each $I$ and $V_{I,+}\subseteq P\subseteq V_{I}$, define the set
\[
E_{I,P} = \sett{x\in \{0,1\}^n}{\AND_{T,j,+}(x)=1\iff (T,j)\in I,\qquad \forall i\in V_I, x_i=1\iff i\in P}.
\]
We note that $E_{I,P}$ forms a partition of $\{0,1\}^n$, and the next claim says that $h_{\vec{y}}$ is constant on each one of these parts.
\begin{claim}\label{claim:constant_on_IP}
    For all $(I,P)$, $h_{\vec{y}}$ is constant on $E_{I,P}$.
\end{claim}
\begin{proof}
    Fixing $(I,P)$, we get that for each $x\in E_{I,P}$, we know that the set of positive ANDs that fire is exactly $I$. We know the value of $x_i$ for each $i\in V_I\setminus V_{I,+}$, so we know precisely which functions $\AND_{T,j}$ evaluate to $1$ on $x$ and which evaluate to $0$. As the function $h_{\vec{y}}$ is a linear combination of products of $\AND_{T,j}$, it follows it is constant on $E_{I,P}$.
\end{proof}

Using Claim~\ref{claim:constant_on_IP}, we choose the value $b(I,P)\in \{0,1\}$ to be the closest to the value of $h_{\vec{y}}$ on $E_{I,P}$. Taking $r = \frac{3m4^d}{\eta\eps}$, we define the decision list $D(x)$ to be as follows: the rules are $(P, b(I,P))$ for all $\card{I}\leq r$, and we order them decreasing lexicographic order of $(\card{I},\card{P})$. We have the following claim asserting that $D(x)$ computes the correct value on any input which is not covered by too many positive ANDs.
\begin{claim}\label{claim:correct_not_too_covered}
    Suppose that $x\in\{0,1\}^n$ is an input such that 
    $\sum\limits_{T\in\mathcal{H}}\sum\limits_{j=1}^{m}\AND_{T,j,+}(x)\leq r$. Then 
    \[
    D(x) = 1_{h_{\vec{y}}(x)\geq 1/2}.
    \]
\end{claim}
\begin{proof}
    Let $I_x = \{(T,j)~|~\AND_{T,j,+}(x)=1\}$, and let $P_x = \sett{i\in V_{I_x}}{x_i=1}$. Then by definition $x\in E_{I_x,P_x}$ and by assumption $\card{I_x}\leq r$, so $D(x)$ contains a rule that leads to the desired value $b(I_x,P_x)$. As $h_{\vec{y}}$ is constant on $E_{I_x,P_x}$, this value is equal to $1_{h_{\vec{y}}(x)\geq 1/2}$. It remains to prove there is no earlier rule that leads to a different value.

    If $(I,P)$ is a different rule that is satisfied by $x$, then we automatically know that $I\subseteq I_x$. Thus, if $I\subsetneq I_x$ then the rule occurs later and we are done. If $I=I_x$ then we must have that $V_{I_x,+}\subseteq P\subseteq V_{I_x}$ and $x_j=1$ for all $j\in P$. In particular, it must be the case that $P\subseteq P_x$, so again the rule is later and we are done.
\end{proof}
The next claim bounds the width of $D$
\begin{claim}\label{claim:width_of_D}
$D$ is a decision list of width $rd = 2^{O(K/\eps)}$.
\end{claim}
\begin{proof}
    For each rule $(I,P)$ in $D$ we have that $\card{P}\leq \card{I}d\leq rd$, and the result follows.
\end{proof}

\subsection{Concluding the Proof}
We now show that $D$ and $f$ are close to each other. Note that
\[
\norm{f-h_{\vec{y}}}_2^2
\leq
3\norm{f-g}_2^2
+
3\norm{g-h}_2^2
+
3\norm{h-h_{\vec{y}}}_2^2
\leq 15\eps,
\]
where we used~\eqref{eq:remove_high_eg}, Claim~\ref{claim:g_h_close} and~\eqref{eq:unionbd}.
It follows from Markov's inequality that with probability $1-60\eps$ over the choice of $x$, we have that $\card{f(x)-h_{\vec{y}}(x)}< \frac{1}{2}$. From~\eqref{eq:unionbd} and Markov's inequality we get that with probability at least $1-\eps$ over the choice of $x$ we have that 
$\sum\limits_{T\in\mathcal{H}}\sum\limits_{j=1}^{m}\AND_{T,j,+}(x)\leq r$, in which case by Claim~\ref{claim:correct_not_too_covered} we have that $D(x) = 1_{h_{\vec{y}}(x)\geq 1/2}$. Combining via the union bound, we get that with probability at least $1-61\eps$ over the choice of $x$ we have that 
$D(x) = 1_{h_{\vec{y}}(x)\geq 1/2}$ and $\card{f(x)-h_{\vec{y}}(x)}< \frac{1}{2}$, and as $f(x),D(x)\in\{0,1\}$ it follows that $f(x)=D(x)$, finishing the proof.
\section{The Monotone Case}
In this section we explain how to adapt the proof from the previous section to prove Theorem~\ref{thm:main_mono}. We use the notations from the previous section, and the only step that changes is the step in Section~\ref{sec:4th_step}.

\subsection{High-level Overview}
The main issue is that while the function $f$ itself is monotone now, it may not be the case that the function $h_{\vec{y}}$, which we used to construct the decision list, is monotone. Thus, the new ingredient in this case is to come up with a new function $h_{\vec{y}}'$ which is close to $h_{\vec{y}}$, is monotone, and is only a function of a collection of ANDs such that for a typical $x$, only a small number of the ANDs evaluate to $1$. The monotonization procedure of~\cite{LM26} achieves this goal, and it amounts to a clever (but somewhat out-of-the-blue) randomized monotonization procedure. We offer a different monotonization procedure, motivated by the switching operator of~\cite{GGLRS}, which we find to be more intuitive.

To motivate this process, let us first consider the case that $p=1/2$, which is not interesting in the context of Conjecture~\ref{conj:friedgut} as it is subsumed by~\cite{FriedgutJunta}, but still serves as good motivation. There is a natural iterative process that given a function $D\colon \{0,1\}^n\to\{0,1\}$ close to some unknown monotone function $D_{\mono}$, turns it to a monotone function $D'$ which is close to $D_{\mono}$. For each coordinate $i\in [n]$, we define the monotonization operator $S_i$ that acts as follows on a function $D$:
\[
S_i D(x) =
\begin{cases}
\min\{D(x), D(x \oplus e_i)\}, & x_i = 0,\\[4pt]
\max\{D(x), D(x \oplus e_i)\}, & x_i = 1.
\end{cases}
\]
It can be shown that $S_i D$ is monotone along direction $i$ (in the sense that flipping the $i$th coordinate from $0$ to $1$ never decreases the value of $S_i D$). Also, it can be shown that if $D$ was already monotone with respect to direction $j$, then $S_i D$ is also monotone with respect to direction $j$. Thus, the function $D' = S_n\circ S_{n-1}\circ\ldots\circ S_1 D$ is monotone, and it can be shown to be close to $D_{\mono}$ (see~\cite[Section 3]{KMSmono}). 

Unfortunately, the monotonization procedure above does not make much sense over $\mu_p^n$ for $p<1/2$. The issue that the operator $S_i D$ is thought of as swapping the values along an edge of the hypercube $((x_i=0,x_{-i}=y),(x_i=1, x_{-i}=y))$ if $D$ receives values $1$, $0$ in it respectively (thus violating monotonicity), but over the $p$-biased cube these points have different measure. This is a problem since it could radically change the expectation of the function: for example, if $p = o(1/n)$ then monotonizing $\bar{x}_1 \land \cdots \land \bar{x}_n \approx 1$ would result in $x_1 \land \cdots \land x_n \approx 0$. To remedy that, we first embed the $p$-biased cube inside a larger space endowed with the uniform measure. More specifically, we use the following definition.
\begin{definition}
Given $D\colon \{0,1\}^n\to\{0,1\}$, where we think of $\{0,1\}^n$ as equipped with the $p$-biased measure, we define the function $D^{\star}\colon [0,1]^n\to\{0,1\}$ as follows. Given an input $u\in [0,1]^n$, we define the point $x\in \{0,1\}^n$ by $x_i = 1_{u_i\geq 1-p}$, and then define $D^{\star}(u) = D(x)$.   
\end{definition}
When $u\in[0,1]^n$ is sampled uniformly, the distribution of $x$ is $p$-biased. Thus, the function $D^{\star}$ is a natural lifting of $D$ to $[0,1]^n$, which is now equipped with the uniform measure. We now define the monotonization operator over $[0,1]^n$
\begin{definition}\label{def:monotonization}
    Given $D^{\star}\colon [0,1]^n\to\{0,1\}$ and $i\in [n]$, define $S_i D^{\star}\colon [0,1]^n\to\{0,1\}$ as follows. For each $w\in [0,1]^{n-1}$, define $\mu_{i,w} = \Expect{a\in [0,1]}{D^{\star}(u_i=a, u_{-i}=w)}$, and define
    \[
    S_i D^{\star}(u_i=a, u_{-i}=w) = 1_{a\geq 1-\mu_{i,w}}
    \]
    if $\mu_{i,w}>0$ and else $S_i D^{\star}(u_i=a, u_{-i}=w)=0$.
\end{definition}
In words, for any fixing of the coordinates beside $i$, we check what is the measure of $1$'s that $D^{\star}$ has, and then define the values of $S_i D^{\star}(\circ, w)$ to be monotone of the same measure (which amounts to being supported exactly on $[1-\mu_{i,w},1]$). 
With Definition~\ref{def:monotonization} in hand and using the notations up to Section~\ref{sec:4th_step}, our proof proceeds in the following steps:
\begin{enumerate}
    \item We first round the function $h_{\vec{y}}$ to a Boolean function $h_{\vec{y},\text{Bool}}$, which is still only a function of the positive AND basis.
    \item We lift the function $h_{\vec{y},\text{Bool}}$ to the function $h_{\vec{y},\text{Bool}}^{\star}\colon [0,1]^{n}\to\{0,1\}$, and present the analog of the AND basis on the space $[0,1]^n$. We then use the switching operator from Definition~\ref{def:monotonization} to monotonize $h_{\vec{y},\text{Bool}}^{\star}$ and explain how the AND basis changes throughout this process.
    \item We project the resulting monotone function back to the space $\{0,1\}^n$, and show that with high probability we get a function close to $f$, which is only a function of an associated positive AND basis.
    \item Finally, we construct a monotone DNF formula using the same ideas as in Section~\ref{sec:4th_step} to finish the proof.
\end{enumerate}
\begin{remark}
    There is an alternative way, staying purely within the Boolean hypercube $\{0,1\}^n$, that describes the result of the combination of the operations of lifting a function from $\{0,1\}^n$ to $[0,1]^n$, applying monotonization in that domain, and then projecting randomly back to $\{0,1\}$. It amounts to going over all coordinates $i=1,\ldots,n$, inspecting all violating edges $(x,x\oplus e_i)$ along it (meaning edges where $x_i=0$ but the value of the function along these endpoints is $(1,0)$) and randomly flipping the value on all of them together to either $(1,1)$ with probability $\frac{1-2p}{1-p}$ or to $(0,1)$ with probability $\frac{p}{1-p}$. This results in a process similar to the one used in~\cite{LM26}. 
\end{remark}

\subsection{The Monotonization Procedure and ANDs}
In this section we discuss functions over $[0,1]^n$, the analog of AND functions in this domain, and how the monotonization procedure affects that. 
\subsubsection{Basic Properties of Monotonization}
The following lemma establishes basic properties of the monotonization operator, analogous to well-known ones on the Boolean hypercube.
\begin{lemma}\label{def:mono_triv}
    Let $F\colon [0,1]^n\to\{0,1\}$ be any function.
    \begin{enumerate}
        \item For all $i\in [n]$, $S_i F$ is monotone in coordinate $i$.
        \item For all distinct $i,j\in [n]$, if $F$ is monotone in coordinate $j$, then $S_i F$ is monotone in direction $j$.
        \item For all $i\in [n]$ and all $G\colon [0,1]^n\to\{0,1\}$ 
        \[
        \Prob{u\in [0,1]^n}{S_i F(u)\neq S_i G(u)}\leq 
        \Prob{u\in [0,1]^n}{F(u)\neq G(u)}.
        \]
    \end{enumerate}
\end{lemma}
\begin{proof}
    The first item is clear by definition. For the second item, note that the function 
    \[
    \mu_{i,w}(F) = \Expect{a\in[0,1]}{F(u_i=a, u_{-i}=w)}
    \]
    is monotone in direction $j$, so by definition we get that $S_i F$ is monotone in direction $j$. For the third item, letting $w\in [0,1]^{n-1}$ be any fixing of coordinates $j\neq i$, we show that
    \[
     \Prob{a\in [0,1]}{S_i F(a,w)\neq S_i G(a,w)}\leq 
        \Prob{a\in [0,1]}{F(a,w)\neq G(a,w)},
    \]
    and then the claim holds by averaging over $w$. Note that the left hand side is equal to 
    $\card{\mu_{i,w}(F)-\mu_{i,w}(G)}$, whereas the right hand side is equal to
    \[
    \Expect{a}{\card{F(a,w)-G(a,w)}}
    \geq 
    \card{\Expect{a}{F(a,w)-G(a,w)}}
    =\card{\mu_{i,w}(F)-\mu_{i,w}(G)},
    \]
    where we used the triangle inequality.
\end{proof}

Define the operator $S_{[n]} = S_n\circ S_{n-1}\circ\cdots\circ S_1$. The following corollary asserts that if $F$ is a monotone function, and it is close to a function $G$, then $S_{[n]}G$ is close to $F$.
\begin{lemma}\label{lem:close_to_mono}
    Let $F\colon [0,1]^n\to\{0,1\}$ be a monotone function, and let $G\colon [0,1]^n\to\{0,1\}$ be a function such that $\Prob{u\in [0,1]^n}{F(u)\neq G(u)}\leq \eps$. 
    Then
    \[
    \Prob{u\in [0,1]^n}{F(u)\neq S_{[n]}G(u)}\leq \eps.
    \]
\end{lemma}
\begin{proof}
    Since $F$ is monotone, $S_{[n]} F = F$, so the left-hand side of the lemma is equal to
    \[\Prob{u\in [0,1]^n}{S_{[n]} F(u)\neq S_{[n]}G(u)}
    \leq \Prob{u\in [0,1]^n}{S_n\circ S_{n-1}\circ\cdots\circ S_1 F(u)\neq S_n\circ S_{n-1}\circ\cdots\circ S_1 G(u)}.
    \]
    Applying the third item of Lemma~\ref{def:mono_triv} $n$ times, the latter probability is at most
    $\Prob{u\in [0,1]^n}{F(u)\neq G(u)}\leq \eps$. 
\end{proof}

\subsubsection{AND Functions on the Solid Cube}
We now define the analogs of AND functions on the solid cube, and the notion of being a function of a collection of ANDs.
\begin{definition}
    A generalized-AND function on $[0,1]^n$ is a function of the form
    \[
    G(u) = \prod\limits_{i\in W} 1_{u_i\in Q_i},
    \]
    where $W\subseteq [n]$ and $Q_1,\dots,Q_n$ are arbitrary measurable subsets of $[0,1]$. We say that $G$ is an AND on coordinate $i\in [n]$ if $Q_i = [1-p,1]$, and say that $G$ is an AND function if $Q_i = [1-p,1]$ for all $i\in W$.
    
    We refer to $\card{W}$ as the width of $G$, and often denote $\mathsf{vars}(G) = W$.
\end{definition}
If $F_1,\ldots,F_N\colon [0,1]^n\to\{0,1\}$ are all generalized-AND functions, and $I\subseteq [N]$, we denote $V_I = \bigcup_{i\in I}\mathsf{vars}(F_i)$.
\begin{definition}
    Let $\mathcal{C} = \{F_1,\ldots,F_N\colon [0,1]^n\to\{0,1\}\}$ be a collection of generalized-AND functions. We say that a function $F\colon [0,1]^n\to\{0,1\}$ is determined by $\mathcal{C}$ if for each $I\subseteq [N]$, the function $F$ is constant on the set
    \[
    \{u\in [0,1]^n~|~F_i(u)=1\iff i\in I\}.
    \]
\end{definition}

\subsubsection{AND function and Monotonization}
Suppose that $F\colon [0,1]^n\to\{0,1\}$ is determined by a collection of AND functions $\mathcal{C}$, and suppose we start running the monotonization procedure described earlier. The following lemma asserts that the new function is determined by a modification of $\mathcal{C}$, and specifies how it works.
\begin{definition}
    Let $F(u) = \prod\limits_{i\in W}1_{u_i\in Q_i}$ be a generalized-AND function, and let $j\in [n]$ be a coordinate on which it is an AND or trivial. We define the function $\pi_j(F)$ as follows:
    if $j\not\in W$, $\pi_j(F)=\{F\}$, and else, $\pi_j(F) = \{\prod\limits_{i\in W \setminus \{j\}} 1_{u_i\in Q_i} \cdot 1_{u_j \in [0,p)]}, \prod\limits_{i\in W \setminus \{j\}} 1_{u_i\in Q_i} \cdot 1_{u_j \in [1-p,1]}\}$.

    If $\mathcal{C} = \set{F_1,\ldots,F_N}$ is a collection of generalized-AND functions, we define the collection $\pi_j(\mathcal{C}) = \bigcup_{i=1}^{N}\pi_j(F_i)$.
\end{definition}
In words, the operator $\pi_j$ replaces a set $Q_j\subseteq [0,1]$ with the two intervals near $0$ and $1$, each of the same length as $Q_j$. The following lemma shows that the operator $\pi_j$ provides the appropriate modification for being determined by a collection of AND functions under the monotonization operator $S_j$. 
\begin{lemma}\label{lem:determined_mono}
    Suppose $\mathcal{C}$ is a collection of generalized-AND functions, that $F\colon [0,1]^n\to\{0,1\}$ is determined by $\mathcal{C}$, and let $j\in [n]$ on which each function in $\mathcal{C}$ is either an AND or trivial. Then the function $S_j F$ is determined by $\pi_j(\mathcal{C})$.
\end{lemma}
\begin{proof}
    Let $w\in [0,1]^{n-1}$ be a fixing of all coordinates except for $j$, and denote $\mu_{j,w} = \Expect{a\in[0,1]}{F(a,w)}$. Then $S_jF(a,w) = 1_{a\geq 1-\mu_{j,w}}$ if $\mu_{j,w} \neq 0$, and $S_jF(a,w) = 0$ if $\mu_{j,w} = 0$. Explicitly, we have
    \[
     S_jF(a,w) =
     \begin{cases}
         F(0,w) \land F(1,w) & \text{if } a \in [0,p), \\
         F(0,w) & \text{if } a \in [p,1-p), \\
         F(0,w) \lor F(1,w) & \text{if } a \in [1-p,1].
     \end{cases}
    \]

    Let $\phi(a) = 0$ if $a \in [p,1-p)$ and $\phi(a) = 1$ otherwise. Given the data $G(a,w)$ for all $G \in \pi_j(\mathcal{C})$ we can determine $F(0,w)$ and $F(\phi(a),w)$, and furthermore, if $F(0,w) \neq F(\phi(a),w)$, whether $a \in [0,p)$ or $a \in [1-p,1]$. These clearly suffice to determine $S_jF(a,w)$.
\end{proof}

We define the average coverage parameter of a collection of generalized-AND functions as follows.
\begin{definition}
    Let $\mathcal{C}$ be a collection of generalized-AND functions. Define
    \[
    \mathsf{AveCov}(\mathcal{C})=\Expect{u\in [0,1]^n}{\sum\limits_{G\in \mathcal{C}}G(u)}.
    \]
\end{definition}

Let $\pi_{[n]}=\pi_{n}\circ\pi_{n-1}\circ\ldots\circ \pi_1$. The following lemma is immediate by iterating Lemma~\ref{lem:determined_mono}.
\begin{lemma}\label{lem:iterate_change_and}
    Let $\mathcal{C}$ be a collection of AND functions of width $w$, and let $F\colon [0,1]^n\to\{0,1\}$ be a function determined by  $\mathcal{C}$. Then $S_{[n]} F$ is determined by the collection $\pi_{[n]}(\mathcal{C})$, and it holds that $\mathsf{AveCov}(\pi_{[n]}(\mathcal{C}))\leq 2^w\mathsf{AveCov}(\mathcal{C})$.
\end{lemma}
\begin{proof}
    The first part is obvious by Lemma~\ref{lem:determined_mono}. For the second part, we note that for any function $G\in\mathcal{C}$ we have that $\sum_{H \in \pi_{[n]}(G)}\Expect{u}{H(u)}\leq 2^w\Expect{u}{G(u)}$.
\end{proof}

By construction, the collection $\pi_{[n]}(\mathcal{C})$ can be written explicitly as follows. We replace each $G \in \mathcal{C}$ by products $\prod_{i \in \mathsf{vars}(G)} 1_{u_i \in Q_i}$, where $Q_i \in \{[0,p), [1-p,1]\}$, in all possible ways.
It will be slightly easier in the sequel instead to consider instead the collection
\[
 \tilde\pi_{[n]}(\mathcal{C}) = \bigcup_{G \in \mathcal{C}} \prod_{i \in \mathsf{vars}(G)} 1_{u_i \in [0,p) \cup [1-p,1]}.
\]

The function $S_{[n]}F$ is no longer determined by $\tilde\pi_{[n]}(\mathcal{C})$. However, it is easy to check that it satisfies the following weaker notion:

\begin{definition}
    Let $\mathcal{C} = \{F_1,\ldots,F_N\colon [0,1]^n\to\{0,1\}\}$ be a collection of generalized-AND functions in which each $Q_i$ is $[0,p) \cup [1-p,1]$. We say that a function $F\colon [0,1]^n\to\{0,1\}$ is pointwise-determined by $\mathcal{C}$ if for each $I\subseteq [N]$ and $v\in [0,1]^n$ such that $\sett{i}{F_i(v)=1}=I$, the function $F$ is constant on the set
    \[
    \{u\in [0,1]^n~|~F_i(u)=1\iff i\in I, u_{V_I}=v_{V_I}\}.
    \]    
\end{definition}

\begin{corollary}\label{cor:iterate_change_and}
Let $\mathcal{C}$ be a collection of AND functions of width $w$, and let $F\colon [0,1]^n \to \{0,1\}$ be a function determined by $\mathcal{C}$. Then $S_{[n]}F$ is pointwise-determined by the collection $\tilde\pi_{[n]}(\mathcal{C})$, and it holds that $\mathsf{AveCov}(\tilde\pi_{[n]}(\mathcal{C}))\leq 2^w\mathsf{AveCov}(\mathcal{C})$.
\end{corollary}

\subsection{The Monotone Approximation}
We now use the tools from the previous two sections to complete the proof of Theorem~\ref{thm:main_mono}. 
We start with $h_{\vec{y}}$ and the collection of AND functions
\begin{align*}
\mathcal{C}
&=\sett{\AND_{T,j,+}(x)}{T\in \mathcal{H},j=1,\ldots,m}\\
&\cup 
\sett{\AND_{T,j,+}(x)x_s}{T\in \mathcal{H},j=1,\ldots,m, x_s\text{ appears in }\AND_{T,j,-}}
\end{align*}
It is easy to see that $h_{\vec{y}}$ is determined by $\mathcal{C}$. Letting $F(x) = 1_{h_{\vec{y}}(x)\geq 1/2}$, we get that $F$ is determined by $\mathcal{C}$ as well as the following fact.
\begin{fact}
    $\norm{F-f}_2^2\leq 60\eps$.
\end{fact}
\begin{proof}
    By~\eqref{eq:remove_high_eg}, Claim~\ref{claim:g_h_close} and~\eqref{eq:unionbd} we have $\norm{f-h_{\vec{y}}}_2^2\leq 15\eps$, hence $\card{f(x)-h_{\vec{y}}(x)}< 1/2$ for all but $60\eps$ fraction of $x$. For any such $x$ we have that $f(x) = F(x)$, and the statement follows.
\end{proof}

\paragraph{Lifting to the solid cube:} take $F^{\star}$ the lift of $F$ to $[0,1]^n$, $f^{\star}$ the lift of $f$ to $[0,1]^n$, and $\mathcal{C}^{\star}$ the lift of $\mathcal{C}$ to $[0,1]^n$. Note that by~\eqref{eq:unionbd},
\[
\mathsf{AveCov}(\mathcal{C}^{\star})
\leq 
\mathsf{AveCov}(\mathcal{C})
\leq 3\frac{m(d+1)4^d}{\eta}.
\]
Take $F^{\star}_{\mono} = S_{[n]}(F^{\star})$ and 
$\mathcal{C}^{\star}_{\mono} = \tilde\pi_{[n]}(\mathcal{C}^{\star})$. Note that $\norm{F^{\star}-f^{\star}}_2^2 = \norm{F-f}_2^2\leq 60\eps$ and $f^{\star}$ is monotone, so from Lemma~\ref{lem:close_to_mono} we get that $\norm{F^{\star}_{\mono} - f^{\star}}_2^2\leq 60\eps$. Also, by Corollary~\ref{cor:iterate_change_and} we get that $F^{\star}_{\mono}$ is pointwise determined by $\mathcal{C}^{\star}_{\mono}$ and that $\mathsf{AveCov}(\mathcal{C}^{\star}_{\mono})\leq 2^d\mathsf{AveCov}(\mathcal{C}^{\star})\leq 3\frac{m(d+1)8^d}{\eta}$.

\paragraph{Going back to the hypercube:} for each $i\in [n]$, independently sample $a_i\in [0,1-p)$ and $b_i\in [1-p,1]$ uniformly, and define $R\colon \{0,1\}^n\to\prod\limits_{i=1}^{n}\{a_i,b_i\}$ by $R(x)_i = a_i$ if $x_i = 0$ and $R(x)_i=b_i$ if $x_i=1$. 
Defining $f_{\mono} = F^{\star}_{\mono}\circ R$ and writing $\mathcal{C}_{\mono}^{\star}=\{A_1,\ldots,A_N\}$, we define  
\[
\mathcal{C}_{\mono} = \sett{A_i\circ R}{i=1,\ldots,N}
\]
and $V_I = \bigcup_{i\in I}\mathsf{vars}(A_i)$ for each $I\subseteq [N]$.
\begin{fact}\label{fact:on_avg}
    The following holds:
    \begin{enumerate}
        \item The function $f_{\mono}$ is always monotone, and for all $I\subseteq [N]$, $P\subseteq V_I$ it is constant on
        \[
        E_{I,P} = \sett{x\in\{0,1\}^n}{A_i\circ R(x) = 1\iff i\in I, \qquad\forall j\in V_I, x_j=1\iff j\in P}.
        \]
        \item The collection $\mathcal{C}_{\mono}$ consists only of positive ANDs of width at most $d$.
        \item $\Expect{\vec{a},\vec{b}}{\norm{f-f_{\mono}}_2^2}\leq 60\eps$.
        \item 
        $\Expect{\vec{a},\vec{b}}{\mathsf{AveCov}(\mathcal{C}_{\mono})}\leq 3\frac{m(d+1)8^d}{\eta}$.
    \end{enumerate}
\end{fact}
\begin{proof}
    For the first statement, the monotonicity is clear since if $x\leq y$, then $R(x)\leq R(y)$, and $F^{\star}_{\mono}$ is monotone. The second part of the first item follows since $F_{\mono}^{\star}$ is pointwise determined by $\mathcal{C}_{\mono}^{\star}$, and for each $x,y\in E_{I,P}$, the value of all of the functions from $\mathcal{C}_{\mono}^{\star}$ and the value of coordinates from $V_I$ is the same on both $R(x),R(y)$.

    For the second item, note that each $F_i\in \mathcal{C}_{\mono}^{\star}$ is a function of the form $\prod\limits_{i\in W} 1_{u_i\in [0,p)\cup [1-p,1]}$, and for each $i$, if we chose $a_i<p$ then $R 1_{u_i\in [0,p)\cup [1-p,1]}\equiv 1$, and else $R 1_{u_i\in [0,p)\cup [1-p,1]} (x)= x_i$. The third item follows as the expectation on the left hand side is equal to $\norm{F^{\star}_{\mono}-f^{\star}}_2^2$. The fourth item follows as the expectation on the left hand side is equal to $\mathsf{AveCov}(\mathcal{C}^{\star}_{\mono})$.
\end{proof}
Combining Fact~\ref{fact:on_avg} with Markov's inequality, we may choose $\vec{a},\vec{b}$ such that $\norm{f-f_{\mono}}_2^2\leq 180\eps$ and $\mathsf{AveCov}(\mathcal{C}_{\mono})\leq 9\frac{m(d+1)8^d}{\eta}$.

\paragraph{Constructing the monotone DNF formula:} denote the value of $f_{\mono}$ on $E_{I,P}$ by $b(I,P)$. Take $r = 9\frac{m(d+1)8^d}{\eta\eps}$ and define the monotone DNF formula
\[
D(x) = \bigvee_{\substack{\card{I}\leq r, P\subseteq V_I\\ b(I,P)=1}}\bigwedge_{j\in P}x_j.
\]
Note that its width is at most $d r = 2^{O(K/\eps)}$.
\begin{claim}
    $D(x)\leq f_{\mono}(x)$ for all $x\in\{0,1\}^n$.
\end{claim}
\begin{proof}
    If $D(x) = 1$, then there is non-empty $E_{I,P}$ with $b(I,P)=1$ such that $x_j=1$ for all $j\in P$. Letting $y$ be the indicator vector of $P$, we have that $x\geq y$ and so $f_{\mono}(x)\geq f_{\mono}(y)=1$, where we used the fact that $y\in E_{I,P}$. It follows that $f_{\mono}(x)=1$.
\end{proof}
We now show that $D$ and $f_{\mono}$ are close.
\begin{claim}\label{claim:almost_done}
    $\Prob{x}{D(x)\neq f_{\mono}(x)}\leq \eps$.
\end{claim}
\begin{proof}
    Note that if $f_{\mono}(x)=1$, then $x\in E_{I,P}$ for some $I,P$, so the only way that $D(x)=0$ is if it only holds for $\card{I}>r$. As $\mathsf{AveCov}(\mathcal{C}_{\mono})\leq \eps r$, the claim follows by Markov's inequality.
\end{proof}
Combining Claim~\ref{claim:almost_done} with the fact that $\norm{f-f_{\mono}}_2^2\leq 180\eps$ gives that $\norm{f-D}_2^2\leq 181\eps$, concluding the proof.

\section{Proof of Tail Theorems}
In this section we prove Theorem~\ref{thm:Bourgain} and Theorem~\ref{thm:Bourgain_mono}. The proof proceeds exactly along the lines the proofs in previous sections, but with modifications. 
Following the notations in Section~\ref{sec:proof_of_main}, we first take $d=k$ and $\eta = 2^{-C\cdot d}\eps^{C}$ for some sufficiently large absolute constant $C>0$ instead of the setting therein, and replace~\eqref{eq:remove_high_eg} with the more immediate
\[
\norm{f-g}_2^2=W^{>k}[f]\leq \eps.
\]
Next, we need to modify Claim~\ref{claim:g_h_close} and its proof, as follows (we use the same notations as therein).
\begin{claim}\label{claim:g_h_close_bourgain}
    Suppose $f\colon (\{0,1\}^n,\mu_p^n)\to\{0,1\}$ satisfies that $W^{\geq k}[f]\leq \frac{\eps}{\sqrt{k}}$. Then
    \[
    \norm{g-h}_2^2\leq O(\eps).
    \]
\end{claim}
Given Claim~\ref{claim:g_h_close_bourgain}, the rest of the proof proceeds exactly in the same way, and we do not repeat it here. We henceforth focus on proving Claim~\ref{claim:g_h_close_bourgain}, and first give an overview of the proof.
\begin{enumerate}
    \item \textbf{Recalling the proof of Claim~\ref{claim:g_h_close}:} we broke the contribution of characters in $g-h$ to buckets $\mathcal{B}_i$, and then upper-bounded the total Fourier mass on each one of them separately. More specifically, we bounded it by 
    \[
    p(1-p)\sum\limits_{i=1}^{n}\inner{\xi_i}{\partial_i f}\leq 
    p(1-p)\sum\limits_{i=1}^{n}\norm{\xi_i}_4\norm{\partial_i f}_{4/3}\leq 
    \left(p(1-p)\right)^{1/4}\sum\limits_{i=1}^{n}\norm{\xi_i}_4
    \Icrd_i[f]^{3/4},
    \]
    where we used H\"older's inequality and the Booleanity of $f$. Applying H\"older's inequality again, it suffices to bound $\sum_{i}^{n}\norm{\xi_i}_4^4$, which we did using global hypercontractivity, and $\sum_{i}^{n} \Icrd_i[f]$, which is bounded by assumption. This last part is what fails for us, as $\sum_{i}^{n} \Icrd_i[f]$ may be very large.

    \item \textbf{Applying random restrictions:} to remedy the total influence issue we use random restrictions and appeal to Bourgain's tail theorem. More precisely, we choose a random restriction $\rho$ that fixes each coordinate to $0$ with probability $1-2p$ and otherwise turns it to an unbiased bit from $\{0,1\}$. Then on average, the weight of $f|_{\rho}$ beyond level $k$ can only decrease, and in particular, it is at most $\frac{\eps}{\sqrt{k}}$, so by Bourgain's theorem, $f|_{\rho}$ is close to a junta $F_{\rho}$. 

    This suggests that instead of bounding $\inner{\xi_i}{\partial_i f}$, we should bound $\inner{\xi_i|_{\rho}}{\partial_i f|_{\rho}}$ and then take average over $\rho$. The advantage here is that intuitively, as $f|_{\rho}$ and $F_{\rho}$ are close, the last inner product should be close to $\inner{\xi_i|_{\rho}}{\partial_i F_{\rho}}$; as the function $F_{\rho}$ is a junta, it has bounded total influence, so the contribution of these terms can be bounded using the previous argument.

    \item \textbf{Bounding the switch from $f|_{\rho}$ to the junta:} the last step in the proof bounds the contribution of $\sum_{i:\rho(i)=*}\inner{\xi_i|_{\rho}}{\partial_i (f|_{\rho}-F_{\rho})}$, which we pay by moving from $f|_{\rho}$ to the junta close to it. For a typical $\rho$, we know that the $2$-norm of $f|_{\rho}-F_{\rho}$ is small, but total influence may be large so the previous argument does not work. Thus, we move the differential operator to the other side and push the sum inside, so that now we wish to bound 
    \[
    -2\inner{\sum\limits_{i:\rho(i)=*}(-1)^{x_i}\xi_i|_{\rho}}{f|_{\rho}-F_{\rho}}.
    \]
    As the Fourier supports of the functions $\chi_{i} \xi_i$ are disjoint (we have multiplied by $\chi_i$ as is was removed in $\xi_i$), we get that they are orthogonal, so the $2$-norm of $\sum_{i=1}^{n}\chi_i \xi_i$ is bounded (it is at most $\norm{g-h}_2/\sqrt{p(1-p)}$). The function on the left-hand side of the inner product above is essentially a random restriction of $\sum_{i=1}^{n}\chi_i \xi_i$ (though some care is needed), so its $2$-norm is typically bounded. Combining this, and the fact that $f|_{\rho}$ and $F_{\rho}$ are typically close, Cauchy--Schwarz gives that the inner product above should be small. 
\end{enumerate}

\subsection{Proof of Claim~\ref{claim:g_h_close_bourgain}}
We start as in the proof of Claim~\ref{claim:g_h_close}, define the sets $\mathcal{B}_i$ and $\xi_i(x) = \frac{1}{\sqrt{p(1-p)}}\sum\limits_{S\in\mathcal{B}_i}\widehat{f}(S)\chi_{S\setminus\{i\}}(x)$ so that
\begin{equation}\label{eq:Bourgain_1}
\norm{g-h}_2^2
=\sum\limits_{\card{S}\leq d, S\not\in \mathcal{S}}\widehat{f}(S)^2=
p(1-p)\sum\limits_{i=1}^{n}\inner{\xi_i}{\partial_i f}
=
p(1-p)\sum\limits_{i=1}^{n}\norm{\xi_i}_2^2.
\end{equation}
Next, we recall a consequence of global hypercontractivity that we will need later on. Using~\eqref{eq:global_hyp} we have $\norm{\xi_i}_4^4\leq 5^{4d}\eta \norm{\xi_i}_2^2$, hence
\begin{equation}\label{eq:Bourgain_2}
p(1-p)\sum\limits_{i=1}^{n}\norm{\xi_i}_4^4
\leq 
5^{4d}\eta p(1-p)\sum\limits_{i=1}^n\norm{\xi_i}_2^2=5^{4d}\eta \norm{g-h}_2^2.    
\end{equation}

We use random restrictions, defined as follows, in order to move from $\mu_p$ to $\mu_{1/2}$.
\begin{definition}
    A random restriction $\rho\in \{0,*\}^n$ is sampled by setting, for each $i$ independently, $\rho_i=0$ with probability $1-2p$, and $\rho_i=*$ otherwise. 
\end{definition}
Here and throughout,  $\rho$ denotes random restriction sampled as above. For a function $F\colon (\{0,1\}^n,\mu_p^{n})\to\mathbb{R}$, we will consider the restricted function $F|_{\rho}\colon (\{0,1\}^{\text{stars}(\rho)},\mu_{1/2}^{\text{stars}(\rho)})\to\mathbb{R}$ defined by fixing the $0$'s of $\rho$ into the corresponding coordinates in $F$. 

Using~\eqref{eq:Bourgain_1} we have
\[
\norm{g-h}_2^2 = 
p(1-p) \sum_{i=1}^n \Expect{\rho}{\inner{\xi_i|_\rho}{\partial_if|_\rho}} =
\frac{1-p}{2}\Expect{\rho}{\sum\limits_{i\in\text{stars}(\rho)}\inner{\xi_i|_{\rho}}{\partial_{i}f|_{\rho}}}.
\]
The second equality holds since the inner product does not depend on the value of $\rho(i)$. Conditioning on $i \in \text{stars}(\rho)$ is balanced by a factor of $\Pr[\rho(i)=*] = 2p$.

We will now need Bourgain's theorem on the unbiased cube, and we use the version from~\cite{KKO}.
\begin{thm}\label{thm:KKO_bourgain}
    If $f\colon (\{0,1\}^n,\mu_{1/2}^n)\to\{0,1\}$ satisfies that $W^{\geq k}[f]\leq \frac{\eps}{\sqrt{k}}$, then it is $O(\eps)$-close to a junta of size $\frac{2^{O(k)}}{\eps^{O(1)}}$.
\end{thm}

Let $\delta_{\rho} = \max(\eps, \sqrt{k}W^{\geq k}[f|_{\rho}])$ if 
$W^{\geq k}[f|_{\rho}]\leq 
\frac{1}{\sqrt{k}}$ and $\delta_{\rho}=1$ otherwise. Then as $\Expect{\rho}{W^{\geq k}[f|_{\rho}]}\leq W^{\geq k}[f]\leq \frac{\eps}{\sqrt{k}}$ it follows that 
\begin{equation}\label{eq:expect_of_delta}
\Expect{\rho}{\delta_{\rho}}=O(\eps).
\end{equation}
By Theorem~\ref{thm:KKO_bourgain} we may find a function $F_{\rho}\colon \{0,1\}^{\text{stars}(\rho)}\to\{0,1\}$ which is a $2^{O(k)}\eps^{-O(1)}$ junta such that $\norm{F_{\rho}-f|_{\rho}}_2^2 = O(\delta_{\rho})$. Using~\eqref{eq:expect_of_delta} we get that $\Expect{\rho}{\norm{F_{\rho}-f|_{\rho}}_2^2}=O(\eps)$, and from the above equality we get that
\begin{equation}\label{eq:split_to_2}
\norm{g-h}_2^2 = \frac{1-p}{2}
\underbrace{\Expect{\rho}{\sum\limits_{i\in\text{stars}(\rho)}\inner{\xi_i|_{\rho}}{\partial_i F_{\rho}}}}_{(\rom{1})}
+
\frac{1-p}{2}
\underbrace{\Expect{\rho}{\sum\limits_{i\in\text{stars}(\rho)}\inner{\xi_i|_{\rho}}{\partial_i (f|_{\rho}-F_{\rho})}}}_{(\rom{2})},
\end{equation}
and we bound each summand separately. 

\paragraph{Bounding $(\rom{1})$:} the point here is that $F_{\rho}$ is a junta, so we have a bound on its total influence. More precisely, using H\"older's inequality twice (first on each inner product, then on the sum) we get that
\[
(\rom{1})
\leq 
\Expect{\rho}{\left(\sum\limits_{i\in\text{stars}(\rho)}\norm{\xi_i|_{\rho}}_4^4\right)^{1/4}
\left(\sum\limits_{i\in\text{stars}(\rho)}\norm{\partial_i F_{\rho}}_{4/3}^{4/3}\right)^{3/4}}.
\]
By the Booleanity of $F_{\rho}$, the second sum is $O(I[F_{\rho};\mu_{1/2}])$, which is at most $2^{O(k)}\eps^{-O(1)}$. Plugging that in and using Jensen's inequality gives that
\[
(\rom{1})\leq \frac{2^{O(k)}}{\eps^{O(1)}}\left(\Expect{\rho}{\sum\limits_{i\in\text{stars}(\rho)}{\norm{\xi_i|_{\rho}}_4^4}}\right)^{1/4}
=
\frac{2^{O(k)}}{\eps^{O(1)}}
\left(2p\sum\limits_{i}\norm{\xi_i}_4^4\right)^{1/4}
\leq \frac{2^{O(k)}}{\eps^{O(1)}}\eta^{1/4},
\]
where we used~\eqref{eq:Bourgain_2} in the last transition.

\paragraph{Bounding $(\rom{2})$:} in this case, we do not know anything about the derivatives of $(f|_{\rho} - F_{\rho})$; instead we know that the $L_2$-norm of this function is small. Thus, we use the fact that 
$\inner{\xi_i|_{\rho}}{\partial_i (f|_{\rho}-F_{\rho})} = 
-2\inner{(-1)^{x_i}\xi_i|_{\rho}}{f|_{\rho}-F_{\rho}}$ to get that
\[
(\rom{2})
= -2\Expect{\rho}{\inner{\sum\limits_{i\in\text{stars}(\rho)}(-1)^{x_i}\xi_i|_{\rho}}{f|_{\rho}-F_{\rho}}}
\leq 
2\sqrt{\Expect{\rho}{\left\|{\sum\limits_{i\in\text{stars}(\rho)}(-1)^{x_i}\xi_i|_{\rho}}\right\|_2^2}}
\sqrt{\Expect{\rho}{\norm{f|_{\rho}-F_{\rho}}_2^2}},
\]
where in the last inequality we applied Cauchy--Schwarz twice. The second term is at most $O(\sqrt{\eps})$. The first term is equal to 
\[
\sqrt{\Expect{\rho}{\left\|{\sum\limits_{i=1}^{n}{1_{\rho_i=*}(-1)^{x_i}\xi_i|_{\rho}}}\right\|_2^2}}
=\sqrt{\Expect{\rho}{\Expect{x\sim \mu_{1/2}^{\text{stars}(\rho)}}{\left(\sum\limits_{i=1}^{n}{1_{\rho_i=*}(-1)^{x_i}\xi_i|_{\rho}}\right)^2}}}.
\]
We expand the square and evaluate diagonal terms and non-diagonal terms separately.
\begin{enumerate}
    \item \textbf{Diagonal terms:} the contribution is equal to 
    $\sqrt{\Expect{\rho}{2p\sum\limits_{i=1}^{n}\norm{\xi_i}_2^2}}
    \leq 2\norm{g-h}_2$, where we used~\eqref{eq:Bourgain_1}.
\item \textbf{Non-diagonal terms:} for $i\neq j$ the contribution of the corresponding term is
\[
\Expect{x\sim\mu_p^n,\rho}{1_{\rho_i=\rho_j=*}(-1)^{x_i+x_j}\xi_i(x)\xi_j(x)}.
\]
Consider the function $\zeta(x_i)= \Expect{\rho_i| x_i}{1_{\rho_i=*}}(-1)^{x_i}$; then $\zeta(1) = -1$, whereas $\zeta(0) = \frac{p}{1-p}$. In particular, $\Expect{x_i\sim \mu_p}{\zeta(x_i)}=0$, so it is equal to $c\chi_i(x_i)$ for some constant $c$. Thus, as conditioned on $x$ the coordinates of $\rho$ are independent, we get that the contribution of term $i,j$ is proportional to
\[
\Expect{x}{\chi_{i}(x_i)\xi_i(x)
\chi_{j}(x_j)\xi_j(x)} = 0,
\]
where in the last equality we used the fact that the Fourier supports of $\chi_i \xi_i$ and $\chi_j\xi_j$ are $\mathcal{B}_i,\mathcal{B}_j$, respectively, which are disjoint, and Fact~\eqref{fact:parseval}.
\end{enumerate}
Overall we get that 
$(\rom{2})\leq O(\sqrt{\eps}\norm{g-h}_2)$.

\paragraph{Combining the bounds:} plugging the bounds on $(\rom{1})$ and $(\rom{2})$ into~\eqref{eq:split_to_2} gives that
\[
\norm{g-h}_2^2\leq \frac{2^{O(k)}}{\eps^{O(1)}}\eta^{1/4}
+O(\sqrt{\eps}\norm{g-h}_2).
\]
It thus follows that either $\norm{g-h}_2^2\leq 2\cdot\frac{2^{O(k)}}{\eps^{O(1)}}\eta^{1/4}$ or $\norm{g-h}_2^2\leq 2\cdot O(\sqrt{\eps}\norm{g-h}_2)$. For sufficiently large $C>0$ (recall that $\eta = 2^{-C \cdot k} \eps^C$), the first case gives $\norm{g-h}_2^2\leq \eps$ directly. As per the second case, simplifying gives $\norm{g-h}_2^2=O(\eps)$, and the proof of Claim~\ref{claim:g_h_close_bourgain} is concluded.

\section{Conclusion}\label{sec:conclusion}
\subsection{Tightness Examples}\label{sec:tightness}
Fix $K\geq 1$ thought of as large, take $n = K2^K$ and partition $[n] = B_1\cup\cdots\cup B_{2^K}$, where each $B_i$ has size $K$. Define the function $f(x) = \bigwedge_{i=1}^{2^K}\left(\bigvee_{j\in B_i}x_j\right)$ and consider $f$ over the uniform hypercube. Elementary computations show that the variance of $f$ is $\Omega(1)$, and that the sensitivity at each $x$ where $f(x)=0$ is at most $K$, and hence $I[f]\leq K$. The following claim finishes the example.
\begin{claim}
    If $D\colon \{0,1\}^n\to\{0,1\}$ is computable by a width $2^{K-10}$ positive decision list, then 
    \[
    \Prob{x}{D(x)\neq f(x)}\geq \Omega(1).
    \]
\end{claim}
\begin{proof}
    Sample $y\in\{0,1\}^n$ uniformly among inputs where $f(y)=1$, and obtain $x$ by choosing $i\in [2^K]$ uniformly, setting $x_{B_i} = 0$, and $x_{B_j} = y_{B_j}$ for all $j\neq i$, so that $f(x)=0$. Consider $D(x)$ and $D(y)$, and consider the first satisfied rule in the computation of $D(y)$, say it read the set of variables $S\subseteq [n]$. By the width assumption on $D$, $\card{S}\leq 2^{K-10}$, and since the $B_j$ are disjoint, $S$ can intersect at most $2^{-10}$ fraction of them. Thus, the probability that $S\cap B_i\neq \emptyset$ is at most $2^{-10}$, so $\Prob{x,y}{D(x)\neq D(y)}\leq 2^{-10}$. It follows that
    \[
    \Prob{x,y}{f(x)=D(x), f(y)=D(y)}
    \leq 2^{-10},
    \]
    so either $\Prob{x}{f(x)\neq D(x)}\geq \frac{1}{2}-2^{-11}$ or $\Prob{y}{f(y)\neq D(y)}\geq \frac{1}{2}-2^{-11}$. In the latter case, since the distribution of $y$ is uniform over $f^{-1}(1)$ which has size $\Omega(2^n)$, we get that $f$ and $D$ are $\Omega(1)$-far. In the former case, the distribution of $x$ is uniform on the point in $\{0,1\}^n$ in which exactly a single block $B_j$ evaluates to $0$. An elementary calculation shows that these points constitute 
    \[
    2^K \frac{1}{2^K}\left(1-\frac{1}{2^K}\right)^{2^{K}-1}
    =\Theta(1)
    \]
    of $\{0,1\}^n$, so we again get that $f$ and $D$ are $\Omega(1)$-far.
\end{proof}
\subsection{Total Influence of Decision Lists}
The fact below shows that a small width decision list always has small total influence, thereby showing that Theorem~\ref{thm:main} gives the correct type of ``proper'' characterization in terms of the type of functions it uses.
\begin{fact}\label{fact:bound_ttf}
    Suppose that $D\colon \{0,1\}^n\to\{0,1\}$ is computable by a width $w$ positive decision list. Then $I[D;\mu_p]\leq 2w$ for all $p\in [0,1]$.
\end{fact}
\begin{proof}
    Note that
    \[
    \Expect{x}{\sum\limits_{i: x_i=1}\card{D(x)-D(x\oplus e_i)}}
    =\sum\limits_{i=1}^{n}\frac{1}{2(1-p)} \Icrd_i[D;\mu_p]
    =\frac{1}{2(1-p)}I[D;\mu_p],
    \]
    and we next upper-bound the left-hand side. Fix $x\in\{0,1\}^n$, let $(S,b)$ be the first rule satisfied by $x$, and let $i$ be such that $x_i=1$. Note that if $i\not\in S$, then the rule $(S,b)$ is also satisfied by $x\oplus e_i$, and because we turn the bit $i$ from $1$ to $0$ no earlier rule is satisfied by $x\oplus e_i$. Thus, $D(x)=D(x\oplus e_i)$ for any such $i$, and thus
    $\sum\limits_{i: x_i=1}\card{D(x)-D(x\oplus e_i)}\leq \card{S}\leq w$. 
    We conclude that $I[D;\mu_p]\leq 2(1-p)w\leq 2w$.
\end{proof}
\bibliographystyle{plain}
\bibliography{ref}

\begin{thebibliography}{10}

\bibitem{Bourgain}
Jean Bourgain.
\newblock On the distribution of the {Fourier} spectrum of {Boolean} functions.
\newblock {\em Israel Journal of Mathematics}, 131:269--276, 2002.

\bibitem{FriedgutJunta}
Ehud Friedgut.
\newblock {Boolean} functions with low average sensitivity depend on few
  coordinates.
\newblock {\em Combinatorica}, 18(1):27--35, 1998.

\bibitem{Friedgut}
Ehud Friedgut.
\newblock Sharp thresholds of graph properties, and the {$k$}-{SAT} problem.
\newblock {\em Journal of the American Mathematical Society}, 12(4):1017--1054,
  1999.
\newblock With an appendix by Jean Bourgain.

\bibitem{GGLRS}
Oded Goldreich, Shafi Goldwasser, Eric~P. Lehman, Dana Ron, and Alex
  Samorodnitsky.
\newblock Testing monotonicity.
\newblock {\em Combinatorica}, 20(3):301--337, 2000.

\bibitem{Hatami}
Hamed Hatami.
\newblock A structure theorem for {Boolean} functions with small total
  influences.
\newblock {\em Annals of Mathematics}, 176(1):509--533, 2012.

\bibitem{KKL}
Jeff Kahn, Gil Kalai, and Nathan Linial.
\newblock The influence of variables on {Boolean} functions.
\newblock In {\em Proceedings of the 29th Annual Symposium on Foundations of
  Computer Science (FOCS)}, pages 68--80. IEEE, 1988.

\bibitem{keevash2023sharp}
Peter Keevash and Noam Lifshitz.
\newblock Sharp hypercontractivity for symmetric groups and its applications.
\newblock {\em arXiv preprint arXiv:2307.15030}, 2023.

\bibitem{KLLMcodes}
Peter Keevash, Noam Lifshitz, Eoin Long, and Dor Minzer.
\newblock Forbidden intersections for codes.
\newblock {\em Journal of the London Mathematical Society}, 108(5):2037--2083,
  2023.

\bibitem{KLLM}
Peter Keevash, Noam Lifshitz, Eoin Long, and Dor Minzer.
\newblock Hypercontractivity for global functions and sharp thresholds.
\newblock {\em Journal of the American Mathematical Society}, 37(1):245--279,
  2024.

\bibitem{KMSmono}
Subhash Khot, Dor Minzer, and Muli Safra.
\newblock On monotonicity testing and boolean isoperimetric-type theorems.
\newblock {\em SIAM Journal on Computing}, 47(6):2238--2276, 2018.

\bibitem{KKO}
Guy Kindler, Naomi Kirshner, and Ryan O'Donnell.
\newblock {Gaussian} noise sensitivity and {Fourier} tails.
\newblock {\em Israel Journal of Mathematics}, 225(1):71--109, 2018.

\bibitem{LM26}
Chenghua Liu and Boning Meng.
\newblock Bounded relative boundary implies narrow {DNF} approximation.
\newblock arXiv preprint, 2026.

\bibitem{ODonnell}
Ryan O'Donnell.
\newblock {\em Analysis of {Boolean} Functions}.
\newblock Cambridge University Press, 2014.
\newblock Corrected arXiv edition released in 2021; see arXiv:2105.10386.

\end{thebibliography}
\end{document}